\documentclass[superscriptaddress,twocolumn,showpacs,pra,longbibliography]{revtex4-1}

\usepackage{amsmath,amssymb,amsfonts}
\usepackage{algorithmic}
\usepackage{graphicx}
\usepackage{textcomp}
\usepackage{xcolor}
\usepackage{braket}
\usepackage{comment}
\usepackage{multirow}

\usepackage{hyperref}
\hypersetup{
    colorlinks=true,       % false: boxed links; true: colored links
    linkcolor=cyan,          % color of internal links
    citecolor=magenta,        % color of links to bibliography
    filecolor=magenta,      % color of file links
    urlcolor=cyan,           % color of external links
    runcolor=cyan
}
\newcommand{\change}[1]{\textcolor{black}{~#1}}

\begin{document}
\title{Towards defending crosstalk-mediated attacks in multi-tenant quantum computing}
\author{Devika Mehra}
\affiliation{Department of Electrical and Computer Engineering, University of Southern California, Los Angeles, California 90089, USA}
\author{Amir Kalev*}
\affiliation{Information Sciences Institute, University of Southern California, Arlington, VA 22203, USA}
\affiliation{Department of Physics and Astronomy, University of Southern California, Los Angeles, California 90089, USA}
\email[Corresponding Author:~]{amirk@isi.edu}

\begin{abstract}
With the increasing demand for quantum hardware, shared and multi-tenant environments have been proposed to optimize resource utilization. However, the multi-tenancy paradigm in quantum computing inherently introduces security threats. This paper examines crosstalk-mediated attacks targeting three-qubit Grover's search algorithm and explores two fundamental mitigation strategies: gate-based dynamical decoupling and the use of a buffer qubit. We evaluate the effectiveness of each method individually and in combination, \textcolor{black}{finding that while both strategies offer some level of attack mitigation, their combined application yields the most significant performance improvement.} Beyond security vulnerabilities, our work also has implications for unintentional circuit interference that can occur when multiple quantum circuits are executed in close proximity.
\end{abstract}

\maketitle
\section{Introduction}
Quantum computing is a rapidly evolving technology with potential applications in many areas of science and technology, notably including many-body physics~\cite{Bauer23}, computational chemistry~\cite{cao2019}, and materials science~\cite{bauer_quantum_2020}. In recent years, there has been a tremendous effort by companies, such as Google, IBM, IonQ, and Rigetti, to develop large quantum information processing units (QPUs) with thousands of physical qubits and at the same time invest resources for developing fault-tolerant quantum computing capabilities, see e.g.,~\cite{ibm_roadmap}. 
In addition to these efforts, some quantum computing companies are offering remote access to their QPUs (quantum cloud computing), servicing hundreds of thousands of users running millions of quantum circuits (jobs) per year~\cite{ibm_users}.

This rapid growth in both QPU size and number of users has encouraged researchers in recent years to develop tools and software that will allow and support  simultaneous access to a single QPU by multiple users~\cite{das_case_2019,niu_multi-programming_2022,liu_qucloud_2024}. 
This multi-tenancy approach aims to improve the throughput of QPUs, for example, by designing compiler tools for QPU multiprogramming, while reducing job submission latency and queue backlogs~\cite{niu_multi-programming_2022}. 

However, multi-tenancy introduces some challenges stemming from having multiple users running quantum circuits in parallel on a single QPU. \textcolor{black}{For example, multi-tenancy can result in degradation of circuit output fidelity due to, e.g., crosstalk effects between qubits used by different users~\cite{niu_multi-programming_2022,upadhyay_share_2024}.} In addition to inadvertently interfering with each other's computations, multi-tenancy raises security concerns. Malicious parties may exploit QPU sharing privileges to interfere with the computation of honest users, and by doing so, potentially compromise the availability and integrity of the computation~\cite{ash2020analysis,saki_qubit_2021,deshpande2022towards,deshpande2023design,harper2024crosstalk,upadhyay_stealthy_2024}.

One such attack that has been considered is the reallocation of hardware priorities. This attack vector could force the victim’s program to use extra SWAP operations, hence increasing their circuit depth and decreasing the computation fidelity~\cite{upadhyay_stealthy_2024}. A more brute-force attack is a crosstalk-mediated attack in which an attacker exploits hardware crosstalk noise channels to compromise the fidelity of the victim’s quantum computation~\cite{ash2020analysis,deshpande2022towards,deshpande2023design,harper2024crosstalk}. This simple attack highlights the need for strong security measures in multi-tenant quantum computing. 

Researchers have suggested a few approaches as a countermeasure to crosstalk-mediated attacks. Ref.~\cite{ash2020analysis} proposed the use of buffer (idle) qubits between users and demonstrated their effectiveness; Ref.~\cite{harper2024crosstalk} developed a machine learning algorithm that aims to optimally map computations onto qubits and showed reduced crosstalk errors using this method; and Refs.~\cite{deshpande2022towards,deshpande2023design} proposed compiling algorithms that scan the input quantum programs to detect malicious patterns. \textcolor{black}{In addition to these approaches, the application of dynamical decoupling (DD) was also suggested as a potentially effective countermeasure for crosstalk-mediated attacks~\cite{ash2020analysis}. However, its effectiveness has not been tested since finding the optimal DD pulse sequence to reduce crosstalk errors is an open question~\cite{ash2020analysis}.}

\textcolor{black}{In this work, we focus on the prospects of DD as a countermeasure for crosstalk-mediated attacks.} Specifically, instead of optimizing DD pulse sequences, we tested the degree to which {\it gate-based} DD (i.e., a gate-level formulation of DD) serves as an effective countermeasure against crosstalk-mediated attacks, whether implemented independently or in conjunction with the buffer qubit strategy.  In the following section, we provide a brief overview of DD and its gate-based formulation. We find that DD gate sequences (specifically, XX and XYXY, see below), while not optimal in terms of their pulse design, are an effective countermeasure for crosstalk-mediated attacks, when implemented by themselves and even more so when combined with buffer qubit strategy. Since generic DD gate sequences, such as XX and XYXY, can be readily implemented through built-in functions in programming languages (such as qiskit), our findings suggest that they should be considered as a useful tool for mitigating potential crosstalk-mediated attacks.

The remainder of the paper is organized as follows. Section~\ref{sec:dd} provides a brief overview of DD, Section~\ref{sec:setup} describes the attacker-victim implementation setup considered in this paper, Section~\ref{sec:results} details the results of our tests, and Section~\ref{sec:conc} offers conclusions.

\section{A brief overview of dynamical decoupling}\label{sec:dd}
DD is a quantum control strategy designed to suppress the impact of environmental disturbances on a quantum system~\cite{viola98,viola_dynamical_1999}. This method achieves its objective by subjecting the system to a rapid succession of carefully timed control pulses that work to diminish unwanted interactions with the environment. Theoretically, the DD framework requires us to implement pulses that are both infinitely fast and infinitely strong to achieve total suppression of noise~\cite{viola98,viola_dynamical_1999}. In practical settings, however, the limitations imposed by finite pulse speeds and amplitudes mean that these ideal effects are only approximated, and neither complete decoupling from the environment is entirely realized. Nevertheless, DD is recognized as one of the simplest and least resource-intensive error mitigation techniques that operates directly at the quantum level to reduce actual errors, as opposed to relying solely on classical post-processing methods for this purpose, see, e.g.,~\cite{biercuk09,lidar2014review,ezzel23}.

\begin{figure}[ht]
\centerline{\includegraphics[width=0.85\linewidth]{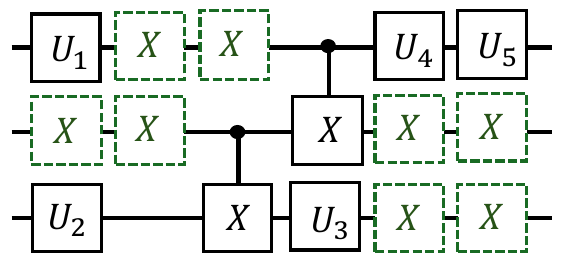}}
\caption{{\bf Example of a gate-based DD.} In this figure two Pauli-$X$ gates are used as a DD sequence within an arbitrary circuit (whose gates are indicated by solid box). The XX sequences do not change the logical operation of the circuit. In addition, in this illustration the DD gates are applied on idle qubits when the time window is sufficiently large to execute them.}
\label{fig:DD}
\end{figure}

In addition to its application as a control strategy in an open quantum system setting, in recent years the DD framework was further developed and studied in the context of quantum cloud computing~\cite{Pokharel18,ezzel23}. In this context, DD is applied as sequences of quantum gates, since remote users generally do not have access to the quantum computing hardware at the pulse level. Moreover, from a computational perspective, the sequence of DD gates must be equivalent to the identity operation, so that the overall computation result is unaffected~\cite{Pokharel18}. Although gate base DD is generally only an approximation to DD, it was found to be effective, in terms of gate overhead, in increasing the fidelity of quantum computation during a variety of tests~\cite{ezzel23}. Two examples of gate-based DD sequences are XX and XYXY, where Pauli-$X$ and Pauli-$Y$ gates are concatenated in single-qubit idle windows to realize a DD sequence~\cite{viola_dynamical_1999}; see Fig.~\ref{fig:DD} for illustration. 
In a recent study~\cite{tripathi2022suppression} XX and XYXY were shown to be effective in mitigating crosstalk errors. Moreover, in Ref.~\cite{Pokharel18} it was demonstrated that applying the DD sequence XYXY achieves a substantial fidelity gain relative to the case where this sequence is not applied. 

Prior studies employing DD as a protocol for crosstalk reduction in quantum cloud computing~\cite{tripathi2022suppression} primarily addressed its application to `spectator` qubits. However, this paper investigates the prospect of utilizing DD on computation qubits to mitigate crosstalk-mediated attacks.

\section{Attacker-victim setup}\label{sec:setup}
\begin{figure*}[htbp]
\centerline{\includegraphics[width=1\linewidth]{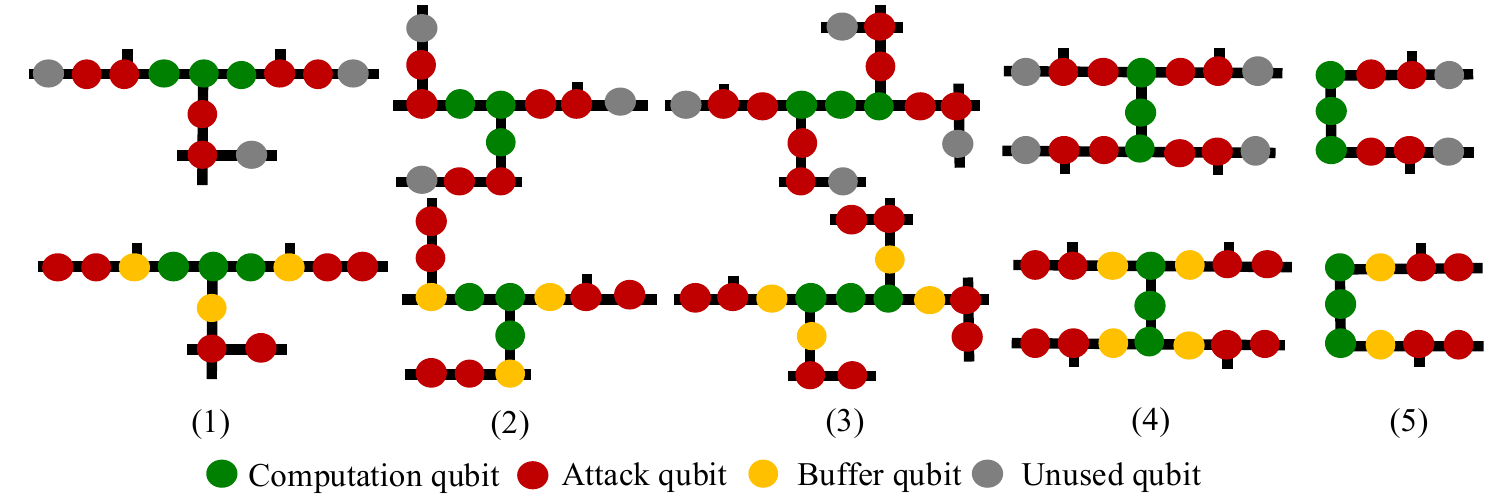}}
\caption{{\bf Attacker-victim  layouts.} This figure illustrates the various attacker-victim connectivity layouts we have tested. The layouts are extracted out from the 127-qubit  ibm\_brisbane QPU. (Top) Attacker qubits, in red, are connected to the victim's (in green) through the QPU connectivity map of ibm\_brisbane. \textcolor{black}{Unused qubits shown in gray are shown in reference to the layouts shown in the bottom figure.} (Bottom) The qubits marked in orange corresponds to buffer qubits, when those are used with or without DD. The specific qubits we used for each layout are given in Tables~\ref{tab:fidelity_differences_geometry_1}-\ref{tab:fidelity_differences_geometry_5} found in Section~\ref{sec:results} and in Appendix~\ref{app:results}.}
    \label{fig:layout}
\end{figure*}

The attack model being studied in this work is one for which the attacker runs its circuit in a close proximity, from a qubit connectivity standpoint, to that of the victim. Whether or not an attacker is present and what qubit assignments it has is not known to the victim. The attacker is presumed to have knowledge of quantum hardware, programming languages applicable to it, and the fundamentals of quantum physics. It is also assumed that publicly available information about the quantum hardware, such as the quality of qubits, gate and channel error rates, and the coupling map along with corresponding strengths, are available to the attacker. Information about crosstalk values can be accessed using idle tomography~\cite{blume2019idle}, which can be useful to decide the optimal attack surfaces around the victim's circuit~\cite{ash2020analysis}. In addition, the attacker may be able to manipulate hardware allocation priorities to reserve access to specific qubits ~\cite{upadhyay_stealthy_2024}. 

Deshpande {\it et al.}~\cite{deshpande2022towards,deshpande2023design} investigated the impact of various attack patterns on two-qubit circuits by implementing, for example, Grover's search, the Deutsch-Jozsa algorithm, and the Bernstein-Vazirani algorithm. Their attacks employed a combination of CNOT and single-qubit gates. They observed that the reduction in the quality of the victim circuit was greater using CNOT gates compared to single-qubit gates. This agrees with the theory reported in~\cite{fang_crosstalk_2022} and the experimental findings~\cite{ash2020analysis} demonstrating that consecutive CNOT operations increase crosstalk errors. \textcolor{black}{On this basis, in this paper we are interested, as a proof of concept, in using repetitive CNOT operations with inserted delays  with different initial states of the control qubit as the primary attack vector as described below in Scenario~3.}

All of the tests reported here were run through cloud access on IBM's 127-qubit QPU ibm\_brisbane. Our tests include five different layouts, in terms of qubit connectivity, and their respective extensions using buffer qubits; see Fig.~\ref{fig:layout}. While chosen ad hoc, the different layouts were used to assess the attack's validity and the countermeasure's effectiveness across various qubit connectivity and placements within the QPU.

\begin{figure}[htbp]
\centerline{\includegraphics[width=1\linewidth]{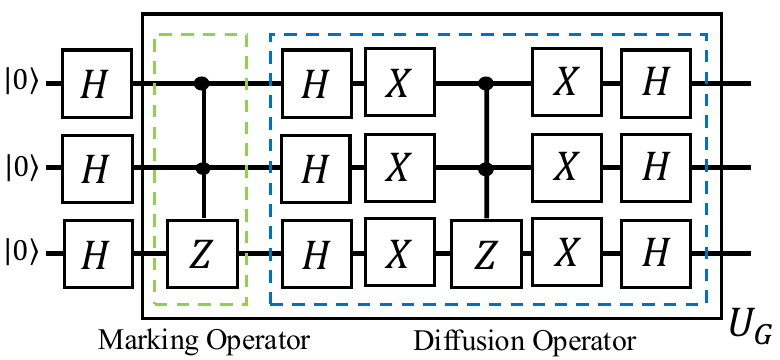}}
    \caption{{\bf Grover search circuit, $U_G$, on three qubits.} The marked item is the bitstring `111'. To obtain high probability of marked item, Grover operator $U_G$ is applied twice.}
    \label{fig:Grover}
\end{figure}

In our tests, the victim runs a Grover search circuit~\cite{grover1996fast} on three qubits; see Fig.~\ref{fig:Grover}. This choice of circuit was made to ensure high-fidelity baseline.
The core of Grover's algorithm involves the repeated application of Grover's operator, $U_G$, which includes a marking oracle and a diffusion operator.
Starting with equal superposition state,
\begin{align}
    \ket{\psi} &= H^{\otimes 3} \ket{000} = \frac{1}{2\sqrt{2}} \sum _{x \epsilon \{0,1\}^3} \ket{x}, 
\end{align}    
the circuit applies two iterations of the Grover's operator $U_G$ (shown in Fig.~\ref{fig:Grover}) to get maximum probability of $\ket{111}$,
\begin{align}
    \ket{\psi _{\text G}} &= U_G ^{2} \ket{\psi} = \frac{11}{8\sqrt{2}} \ket{111} - \frac{1}{8\sqrt{2}}\sum _{x \epsilon \{0,1\}^3/ \{111\}} \ket{x}.
\end{align}  
This gives the probability of getting the marked bitstring $111$ to be $\frac{121}{128}\approx0.945$.

\begin{figure}[htbp]    
    \centerline{\includegraphics[width=1\linewidth]{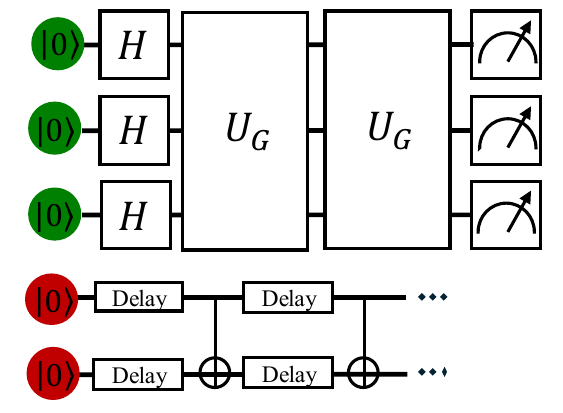}}
    \caption{{\bf Scenario 3: Attack without Mitigation.} In this scenario, the victim implements Grover’s search circuit, cf. Fig.~\ref{fig:Grover}, while the attack qubits undergo a sequence of CNOT operations. The figure illustrates a single pair of attacker qubits; however, in the actual implementation, the same CNOT sequence is applied as an active attack vector to all pairs of attacking qubits surrounding the victim's circuit, following the attack sites specified in Fig.~\ref{fig:layout}. In addition, three attack vectors has been tested, where control qubits are initialized to $\ket0$ (shown in the figure), $\ket{1}$ and $\ket+$. The target qubit is initialized to $\ket0$ in all tested attack vectors. }
    \label{fig:with-attack}
\end{figure}

\begin{figure}[htbp]    
    \centerline{\includegraphics[width=0.85\linewidth]{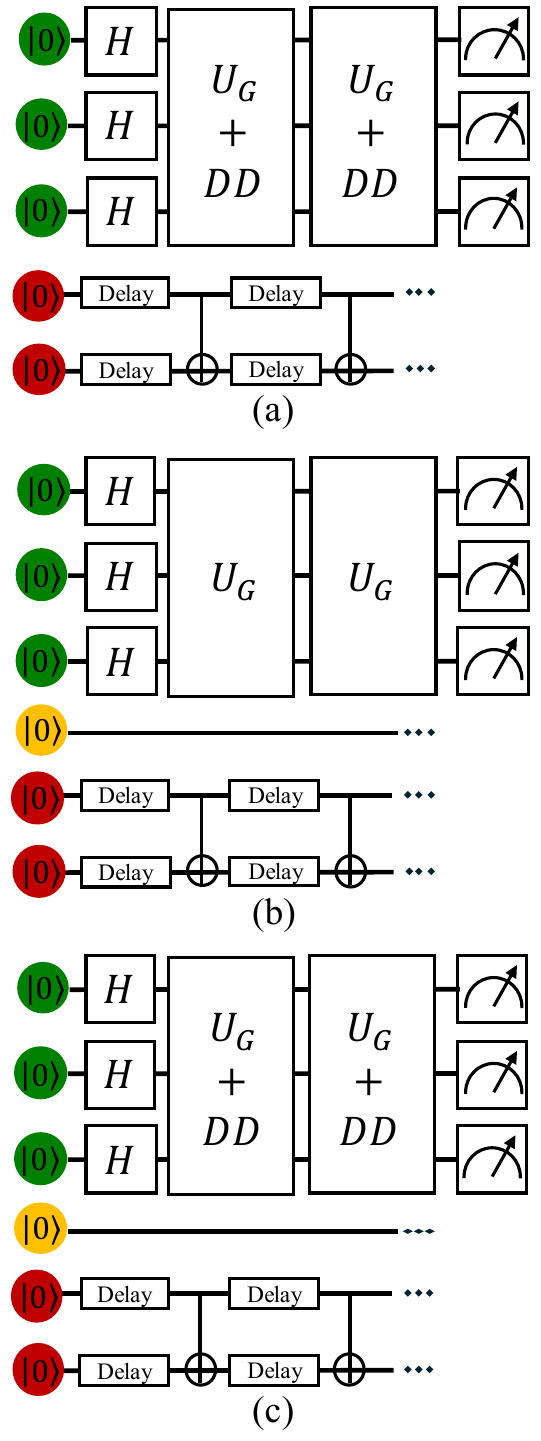}}
    \caption{{\bf Scenario 4: Attack with Mitigation.} We have compared three attack mitigation schemes. (a) Application of DD schemes: The DD sequences are implemented on the victim's circuit (green qubits in Fig.~\ref{fig:layout}), either XYXY or XX sequence. (b) Having a buffer qubit, separating the victim's circuit from that of the attacker. (c) Application of DD (XX or XYXY) on victim circuit with single buffer qubit.}
    \label{fig:with-attack-mit}
\end{figure}

\begin{figure}[htbp]
\centerline{\includegraphics[width=1\linewidth]{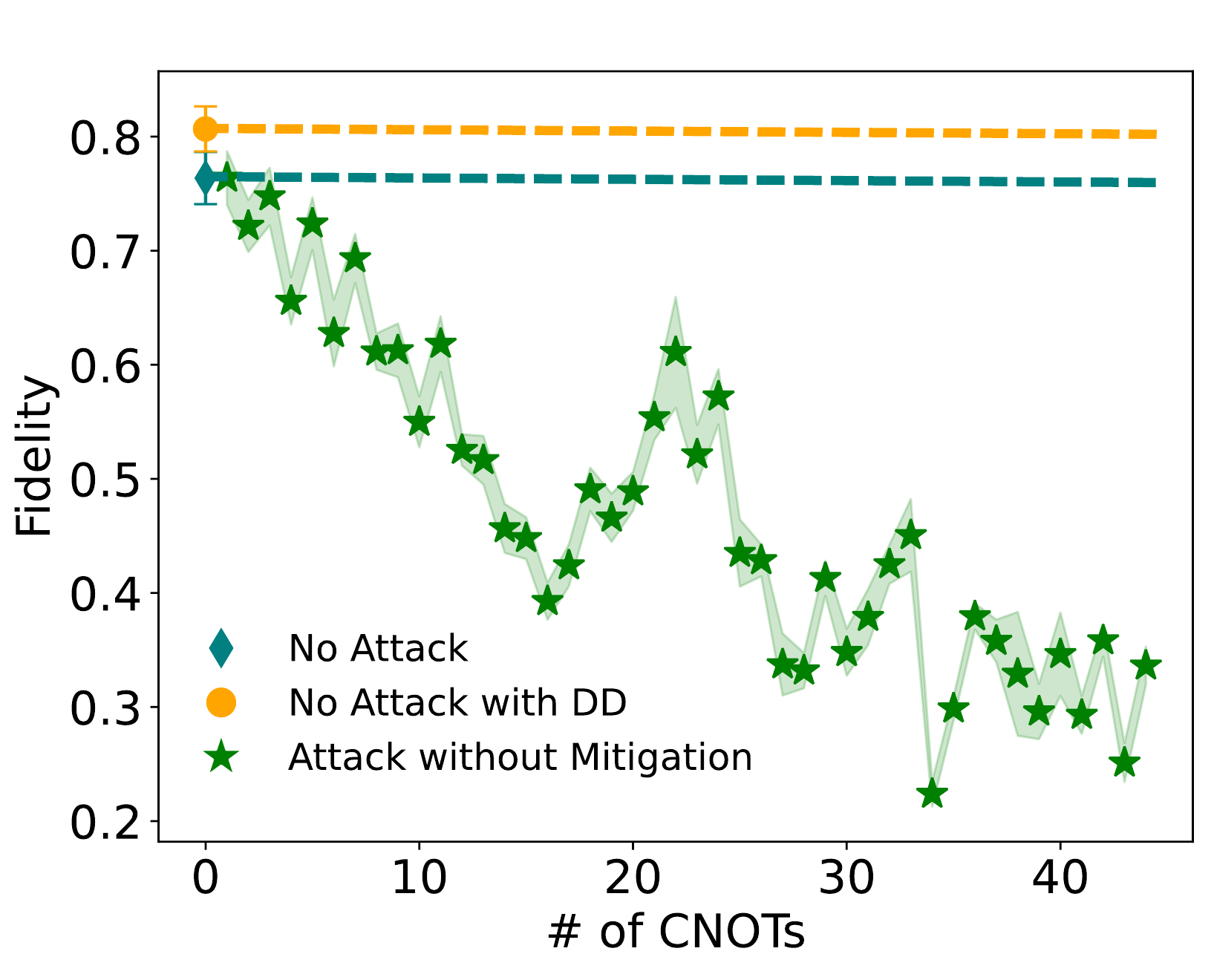}}
    \caption{{\bf Fidelity in Scenarios 1, 2 and 3.} We plot the tested fidelity $F_G$ to the ideal state as a function of the number of CNOT operations on the neighboring attack qubits. This data was observed using layout (1) of Fig.~\ref{fig:layout}. The No Attack with DD data corresponds to applying XYXY sequence, and the attack corresponds to initializing the attack qubits to the $\ket0$  state.  Similar trends were observed with other tests as well (see Fig.~\ref{fig:main_results} and Appendix~\ref{app:results}).}
    \label{fig:prob111}
\end{figure}

Given this attacker-victim setup, we have considered four different scenarios:

\noindent{\textbf{\textit{Scenario 1: No Attack.}}} In this scenario,  Grover's search circuit is run on the three computation qubits (labeled green in Fig.~\ref{fig:layout}). In this scenario, no attack circuit is being executed.   

\noindent{\textbf{\textit{Scenario 2: No Attack with DD.}}}
To establish a reference for the application of DD sequences, in this scenario,  DD gate sequences (either XX or XYXY)  are applied in idle time windows of computation qubits while running Grover's search algorithm. These windows can be found after the transpilation of the control-control-$Z$ ($CCZ$) gate to the native basis gates of ibm\_brisbane QPU. In practice, we used the built-in qiskit module \textit{PadDynamicalDecoupling} to automatically schedule DD sequences. Similar to Scenario~1, no attack is executed in this scenario. 

\noindent{\textbf{\textit{Scenario 3: Attack without Mitigation.}}}
In this scenario, Grover's search circuit is executed similarly to Scenario~1, but here an attack takes place by executing a sequence of CNOT gates on the adjacent pairs of attack qubits surrounding the victim circuit (red qubits in Fig.~\ref{fig:layout}). Three different attack vectors were tested that correspond to three different initial states for the attack qubits (\(\ket{0}\), \(\ket{1}\), and \(\ket{+}\) for the control qubit and \(\ket{0}\) for target qubit).
We perform tests for each attack vector with an increasing number of CNOT gates, up to a total of 45. The CNOT gates are evenly distributed across the duration of Grover's search execution. To avoid nullification of pairs of CNOT gates by the optimization cycle of the qiskit transpiler, delays are inserted between the CNOT gates, and the entire sequence is implemented using the {\it Schedule} and {\it timeline\_drawer} features of the IBM qiskit library. The condensed circuit for this scenario is illustrated in Fig.~\ref{fig:with-attack}.

\noindent{\textbf{\textit{Scenario 4: Attack with Mitigation.}}}
The last scenario, illustrated in Fig.~\ref{fig:with-attack-mit}, includes all the components of the Scenario~3, in addition to including mitigation measures. We tested and compared the effectiveness of application of DD (XX or XYXY), including a qubit buffer, and combination of the two, as means of protecting the victim's circuit.  Importantly, the implementation of DD sequences in this scenario is executed exactly the same manner as in Scenario~2. \textcolor{black}{We note while our study treats mitigation as a user-level choice for experimental clarity, quantum computing providers could implement these measures automatically during qubit allocation.}

\section{Tests and Results}\label{sec:results}

\begin{figure*}[htbp]    
    \centerline{\includegraphics[width=1\linewidth]{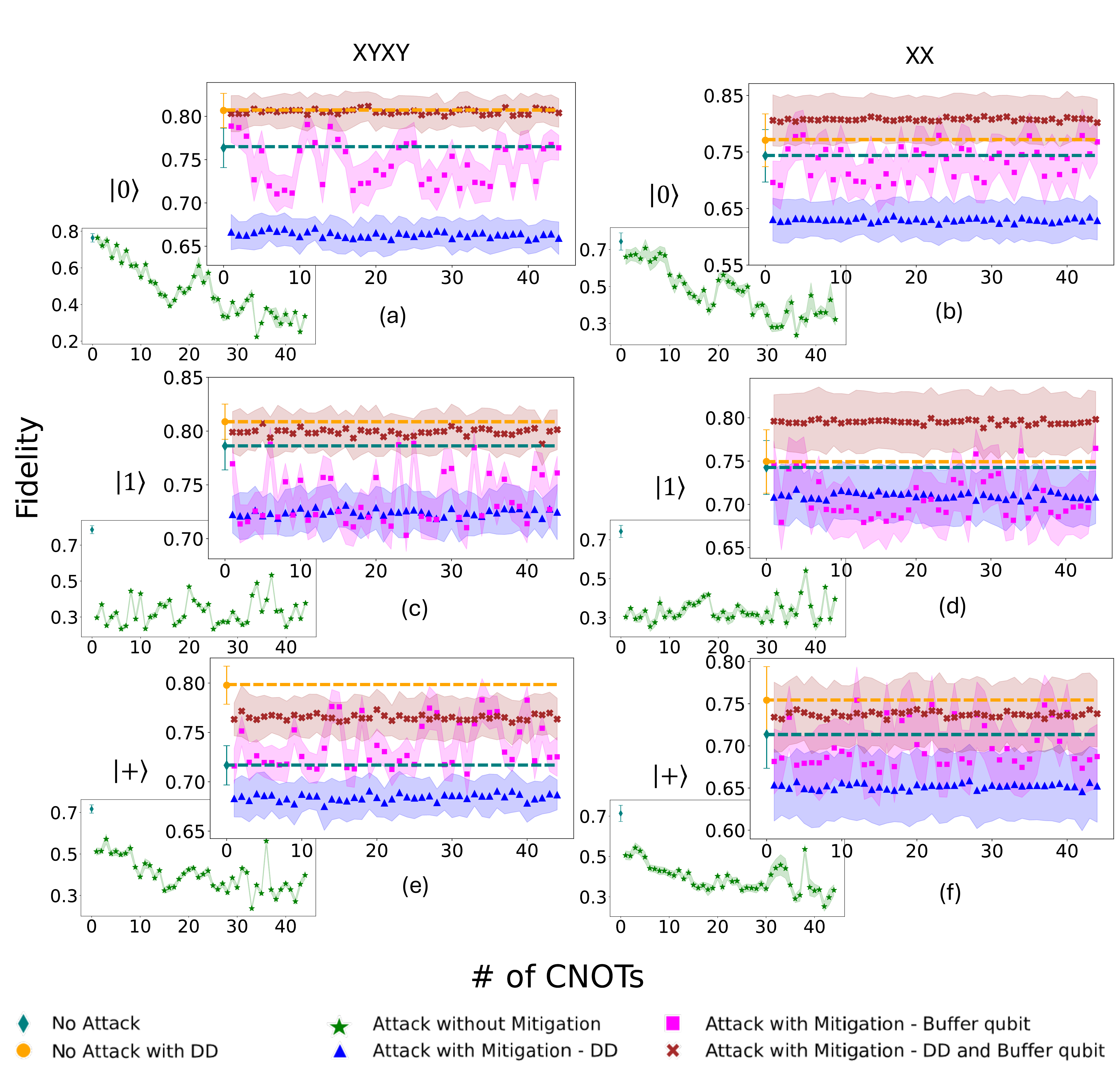}}
    \caption{{\bf Test results for layout (1).} This figure summarizes our results for layout (1) (note the different y-axis range for each subplot). Each data point represents an average over 20 independent tests while the shaded regions indicate the standard deviation across these 20 tests. In all cases, the attacker's control qubit is positioned closest to the victim circuit, with varying initial states. In plots (a) and (b), the attacker starts in $\ket{0}$; in (c) and (d), in $\ket{1}$; and in (e) and (f), in $\ket{+}$. The XYXY DD sequence is applied in (a), (c), and (e), while the XX sequence is used in (b), (d), and (f).
    To induce crosstalk, a series of CNOT gates is incrementally applied to three attack sites in the layout shown in Fig.~\ref{fig:layout}, layout (1). The resulting fidelity degradation as the number of CNOT gates increases (Attack without Mitigation) was observed consistently across all of the tests. When a DD sequence is applied, either XX or XYXY (Attack with Mitigation - DD), the fidelity improves with reduced variance. Alternatively, introducing a buffer qubit (Attack with Mitigation - Buffer qubit), suppresses the attack’s impact while maintaining variance similar to the No Attack scenario. Finally, when both DD and a buffer qubit are applied together (Attack with Mitigation - DD and Buffer qubit), the results closely match those obtained when DD is applied in the absence of an attack (No Attack with DD), demonstrating the combined mitigation strategies' effectiveness.}
\label{fig:main_results}
\end{figure*}

Our tests, as we now describe, were carried out on the 127-qubit ibm\_brisbane QPU~\cite{ibm_quantum}. The basis gate set of the ibm\_brisbane at the time of execution included {ECR (Echoed Cross-Resonance), ID (Identity), RZ (Rotation around the $z$ axis), SX (square of Pauli-$X$), and Pauli-$X$} gates and the device has a CLOPS (Circuit Layer Operations Per Second) of 180K. Additional calibration details can be found in the appendix~\ref{app:hardware} as well as in~\cite{devika_git}. 
The tests were divided into five batches corresponding to the five layouts in Fig.~\ref{fig:layout}, (1) through (5). In each batch, we tested the four scenarios listed above. The specific qubits on which we run the tests are given in Tables~\ref{tab:fidelity_differences_geometry_1}-\ref{tab:fidelity_differences_geometry_5}, (see below and in Appendix~\ref{app:results}) and their performance parameters are given in Table~\ref{tab:qubit_performance} in Appendix~\ref{app:hardware} and in ~\cite{devika_git}. For statistical analysis, each test was repeated twenty times for the layout (1) of Fig.~\ref{fig:layout}, and ten times for each of the remaining layouts. The corresponding quantum circuits were executed using 4096 measurement shots. The distribution of the observed outcomes is used to calculate its fidelity, denoted by $F_{\rm G}$, with the ideal state $\ket{\psi_{\rm G}}$, using \(\it{hellinger\_fidelity}\) from the \(\it{quantum\_info.analysis}\) module of qiskit library.

\subsection{Results}
In this section, we present in detail the results obtained in the tests performed on the layout (1) of Fig.~\ref{fig:layout}. Qualitatively, similar results were observed for other layouts and can be found in appendix~\ref{app:results}. {We note however, that in some test cases that are reported in the Appendix,  little qualitative difference was found between the different mitigation schemes.} The raw data for all tests reported in this paper, as well as the qiskit code developed,is given in~\cite{devika_git}.

To start with, in Fig.~\ref{fig:prob111} the fidelity (averaged over 20 tests),  $F_{\rm G}$, observed in Scenarios 1, 2, and 3 as a function of the number of CNOTs applied in the third scenario, where the initial state for the attack qubits is the all-zero state.  Note that in Scenarios 1 and 2 there is no attack, hence no CNOT gates are applied. 
The results presented in this figure emphasize that even when the attack qubits are initialized to the all-zero state (i.e., when the CNOT gates remain inactive), a significant degradation in the victim's ability to detect the marked item is observed in Scenario~3. This effect of fidelity degradation, previously reported in the literature~\cite{fang_crosstalk_2022,ash2020analysis}, arises from an unintended activation of a ZZ coupling between attacker and victim qubits, when a CNOT-type coupling is present in attacker qubits. This phenomenon, which was observed consistently across our tests, is the basis of crosstalk-mediated attack.  

Next, we compare the effectiveness of different countermeasures to suppress the attack's effect and test their performance compared to the No Attack scenarios (Scenarios 1 and 2). The results are summarized in Fig.~\ref{fig:main_results}. Similarly to Fig.~\ref{fig:prob111}, we plot fidelity $F_{\rm G}$  as a function of the number of CNOT gates applied in Scenario 3. The shaded regions represent the standard deviation of the observed data. 
The top, middle, and bottom rows of Fig.~\ref{fig:main_results} correspond to the three attack vectors where the control qubits of the attacker are initialized to $\ket0$, $\ket1$, and $\ket+$, respectively.
The left and right columns of the figure show the results when in Scenario 4 the DD sequences (when applied) correspond to the XYXY and XX sequences, respectively. When DD was applied, the increase in the circuit depth was \(4\%\) for the XX sequence and \(15\%\) for the XYXY sequence~\cite{devika_git}.

The results highlight a few general characteristics. First, the combination of the two countermeasures, DD and single-qubit buffer, provides the best overall performance, in terms of average fidelity and its fluctuation (or stability) as a function of the number of CNOT gates applied. When both countermeasures are applied, the average fidelity is generally restored to the level of the No Attack scenario, at least, and at few instances exceeds the fidelity level of the No Attack with DD scenario. This underscores the cumulative effect of these two countermeasures in effectively mitigating the attack's impact. Second, we find that in all of our tests the fluctuation in the average fidelity as a function of the number of CNOTs is substantially lower in the case where DD is applied compared to the case where a buffer qubit is introduced. {We believe these fluctuations in the experimentally observed values come from a combination of how the device is calibrated and how  CNOT gates are implemented (in active or idle mode). We plan to study  this phenomena more closely in the future.} Nevertheless, the latter typically outperforms the former in terms of the value of the average fidelity, per CNOT count. In addition, while the average fidelity when DD is applied is consistently below the No Attack fidelity level (and below the No Attack with DD fidelity level), the average fidelity observed when a buffer qubit countermeasure is applied can exceed the corresponding No Attack level.  

\begin{figure}[t!]
    \centering \includegraphics[width=0.95\columnwidth]{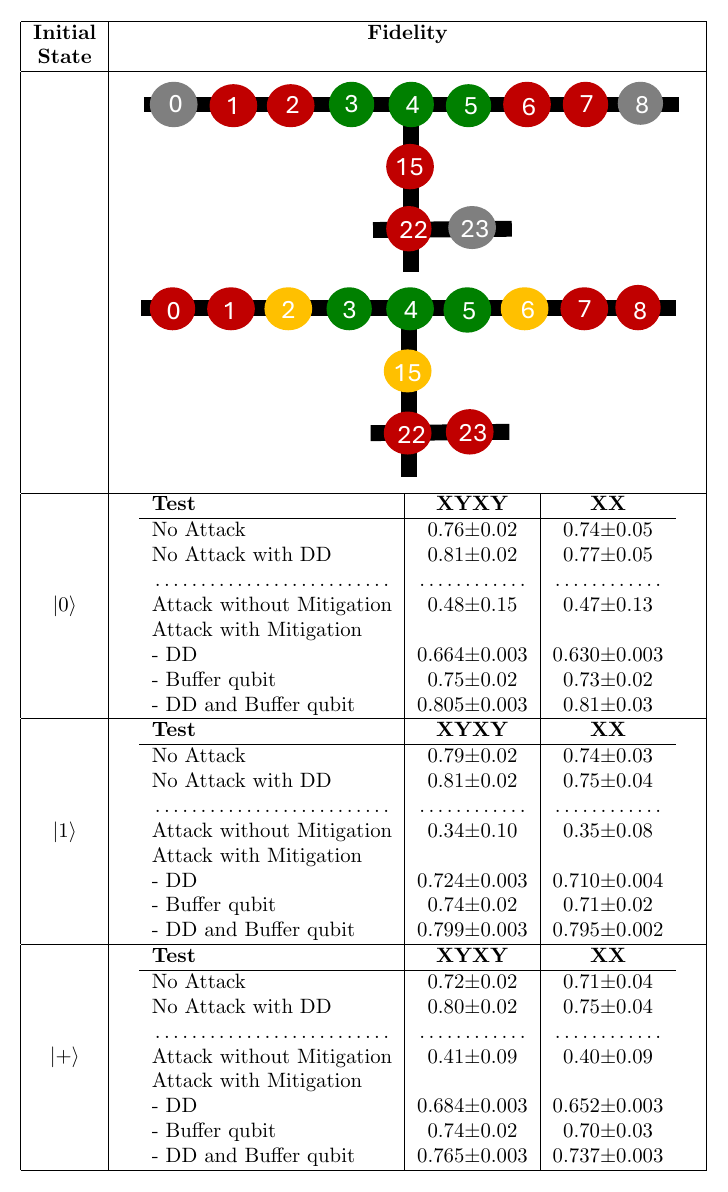}
\caption{{\bf Average fidelity for layout (1)}. This table summarizes the data presented in Fig.~\ref{fig:main_results}. {In the No Attack scenarios the average fidelity and its associated variance  is based on 20 experimental realizations. In the Attack with Mitigation scenarios, the fidelity represents an average over CNOT counts, from 1 to 45, each of which is by itself an average over 20 experimental realizations}.  The figure in table shows the specific qubits within ibm\_brisbane used in these tests.}
\label{tab:fidelity_differences_geometry_1}
\end{figure}

To further highlight the trends found in our tests, we summarize them in Table~\ref{tab:fidelity_differences_geometry_1}, where the results in Scenario 3 and 4 (given in Fig.~\ref{fig:main_results}) are averaged over tests corresponding to different number of CNOTs. The results show that, while neither consistently restores the fidelity to the level of No Attack when applied individually, the DD and buffer qubit might be an effective tool to protect the victim's circuit from potential crosstalk-mediated attacks. Therefore, the choice of mitigation strategy may depend on the user's feasibility space (number of qubits) and \change{circuit depth} constraints. If space is a limited resource, adding a DD sequence may serve as a viable mitigation approach, while if time is the limiting factor, incorporating a buffer qubit is an effective alternative. 

Nevertheless, following the results in Table~\ref{tab:fidelity_differences_geometry_1}, on average, combining the two mitigation schemes, the DD sequences (such as XX or XYXY) together with a single buffer qubit, is an effective and efficient scheme to mitigate against potential crosstalk-mediated attacks. In almost all of the tests the averaged fidelity of the combined countermeasure exceeds that of the Attack without Mitigation and that of the No Attack scenarios.

\section{Conclusion}\label{sec:conc}
In this work, we evaluated the effectiveness of gate-based DD sequences (XYXY and XX) and a buffer qubit in protecting the 3-qubit Grover search circuit from crosstalk-mediated attacks. \textcolor{black}{As a proof of concept we modeled the attack as a simple series of  CNOT gates interleaved with delay operation which was shown to be an effective attack vector in prior works~\cite{ash2020analysis}.} Our findings indicate that each strategy, when applied alone, can effectively mitigate the attack, with some drawbacks. Although the applied DD sequences were generally not able to recover the fidelity in the presence of an attack to that of the No Attack scenarios, it often led to stable fidelity level exceeding that of the Attack without Mitigation scenario. In contrast, while adding a single buffer qubit was found to be a good strategy to mitigate the attack in terms of fidelity, it fell short in terms of the fluctuation of the latter as a function of the number of CNOT gates in the attack. The combination of the two strategies yielded the ``best of both worlds,'' i.e., the consistent fidelity level exceeded the No Attack level and for some tests the No Attack with the DD fidelity level. Although we studied these mitigation strategies in the context of a potential crosstalk-mediated attack, since they introduce a reasonable overhead in terms of qubits and gates, our results suggest that these mitigation techniques may also be used to protect circuits from unintended interference of nearby executions.

\section*{Acknowledgment}
DM would like to express gratitude to Mr. Vinay Tripathi for his invaluable discussions, particularly in the area of error mitigation through DD. The authors thank the three anonymous referees for their insightful comments and constructive criticism, which have been instrumental in refining our claims and enhancing the overall results and presentation of this article. This research was conducted using IBM Quantum Systems provided through USC’s IBM Quantum Innovation Center. 
\bibliography{biblio}

\clearpage
\appendix
\begin{widetext}
\section{Additional Tests}\label{app:results}
This appendix contains all the results we obtained for layouts (2)-(5). The structure of the figures and tables given here follows that of those in the main text.

\begin{figure}[h]
    \centering
    \includegraphics[width=1\linewidth]{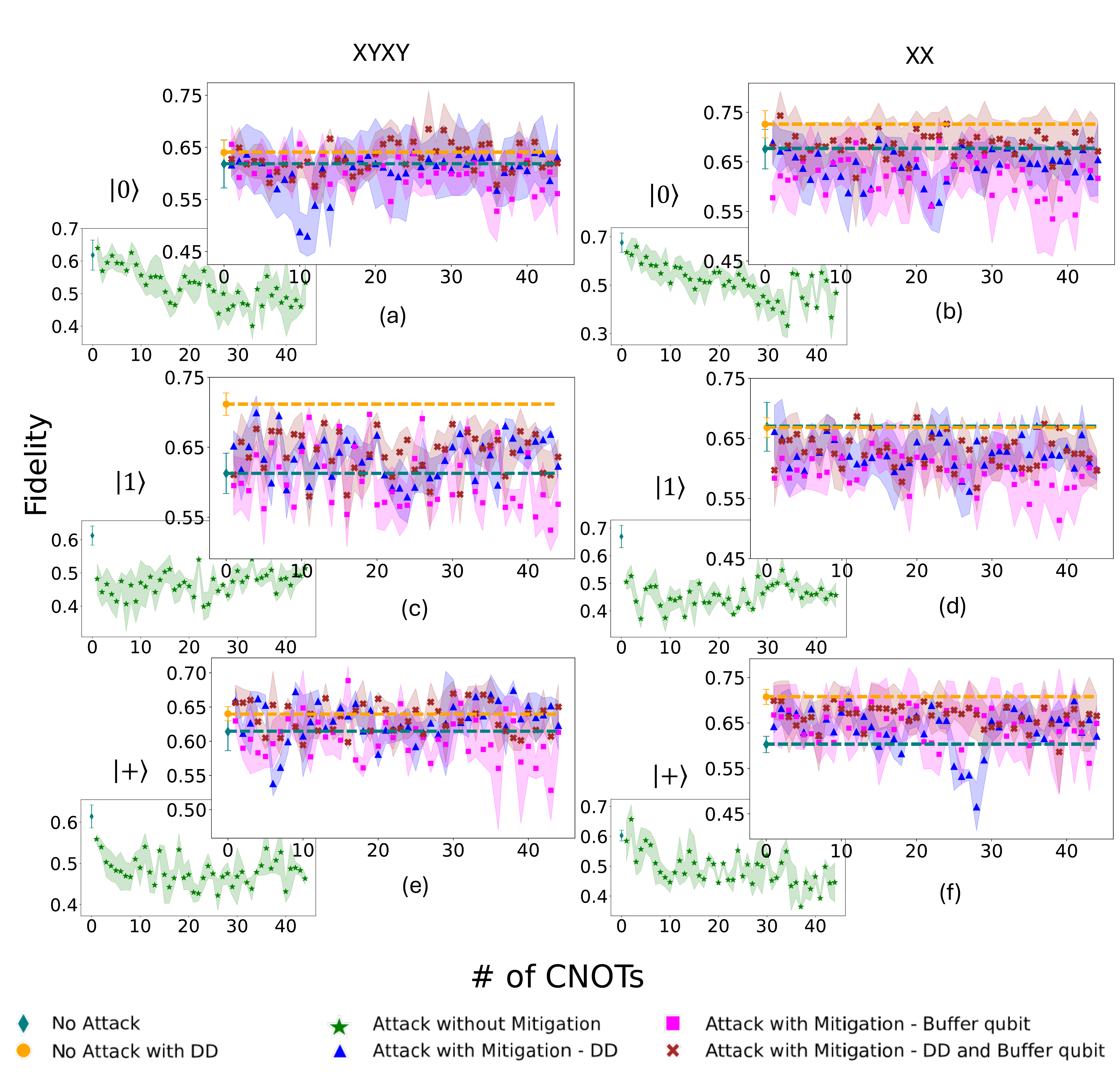}
    \caption{{\bf Test results for layout (2).}}
\end{figure}

\begin{figure}[h]
    \centering
    \includegraphics[width=1\linewidth]{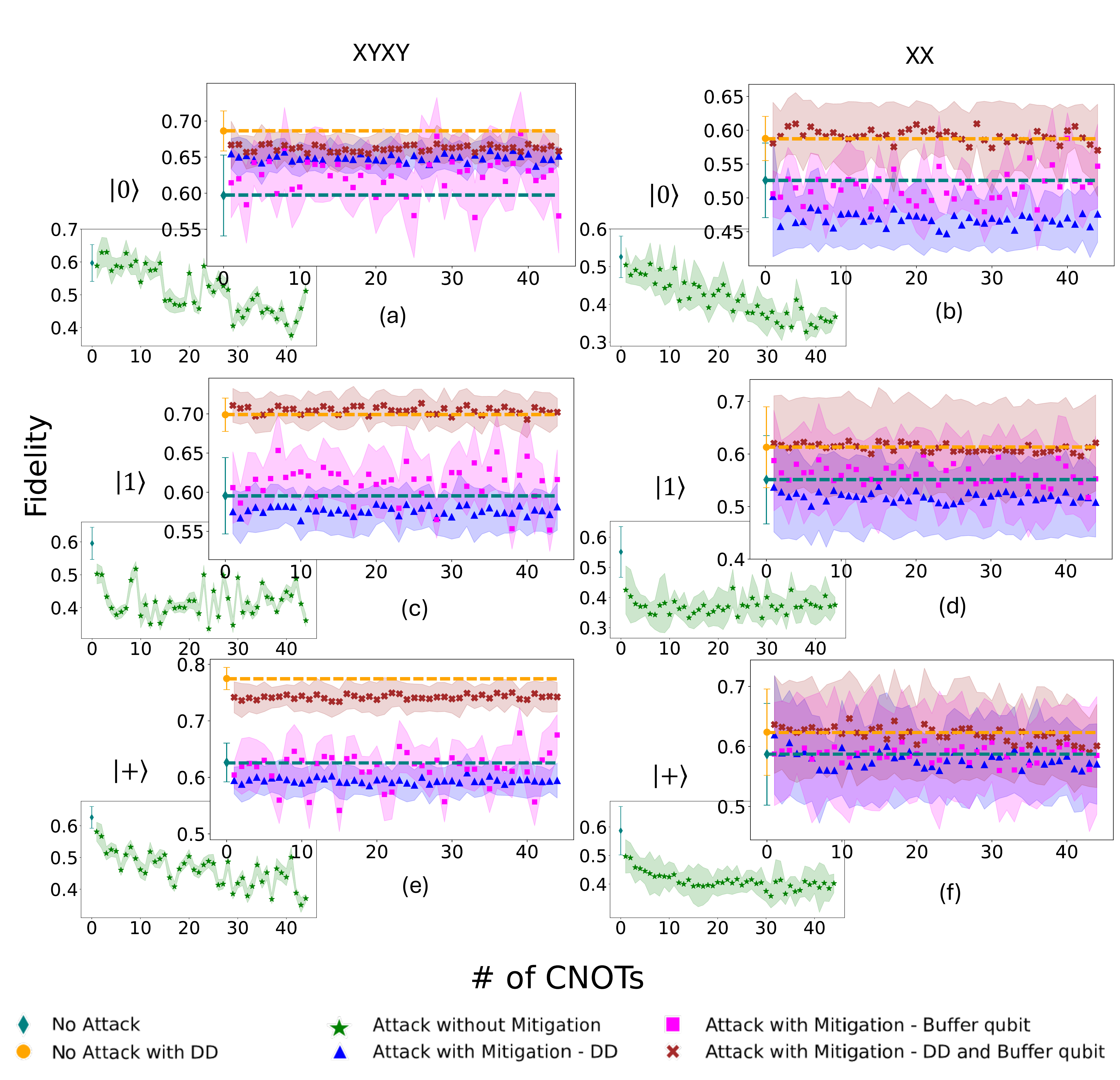}
    \caption{{\bf Test results for layout (3).}}
\end{figure}

\begin{figure}[h]
    \centering
    \includegraphics[width=1\linewidth]{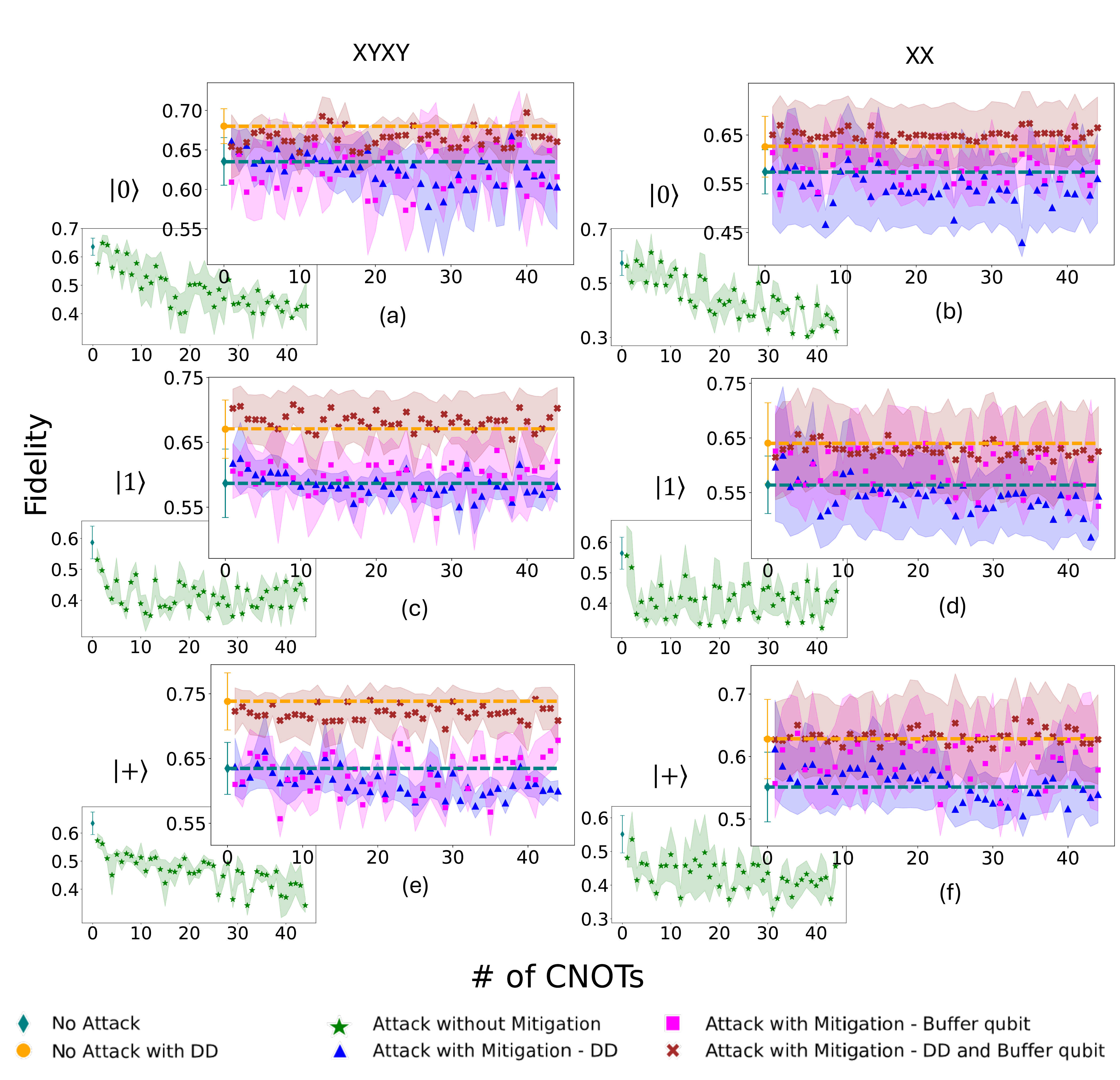}
    \caption{{\bf Test results for layout (4).}}
\end{figure}

\begin{figure}[h]
    \centering
    \includegraphics[width=1\linewidth]{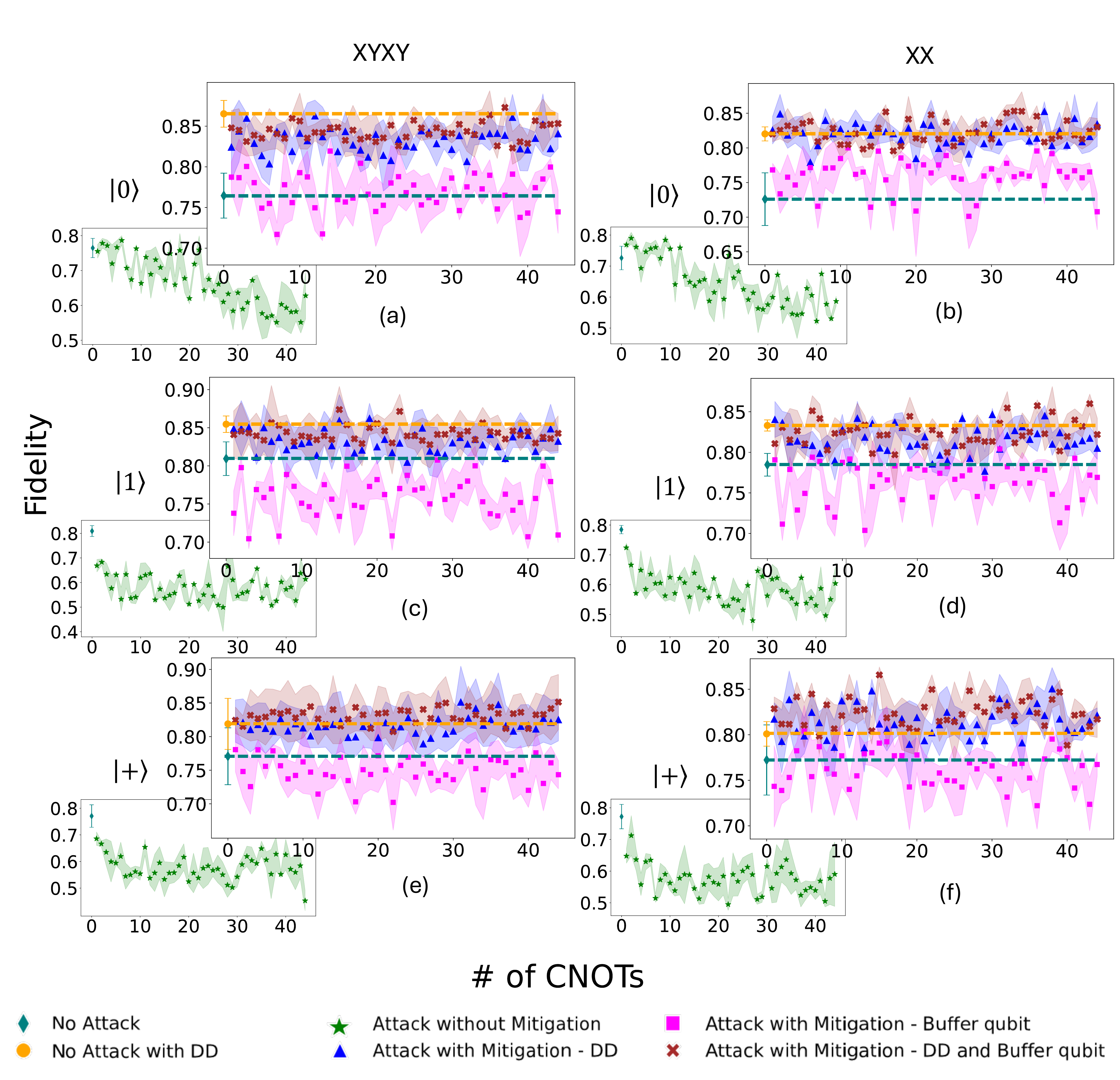}
    \caption{{\bf Test results for layout (5).}}
\end{figure}
\end{widetext}

\clearpage

\begin{figure}[t!]
    \centering \includegraphics[width=0.95\columnwidth]{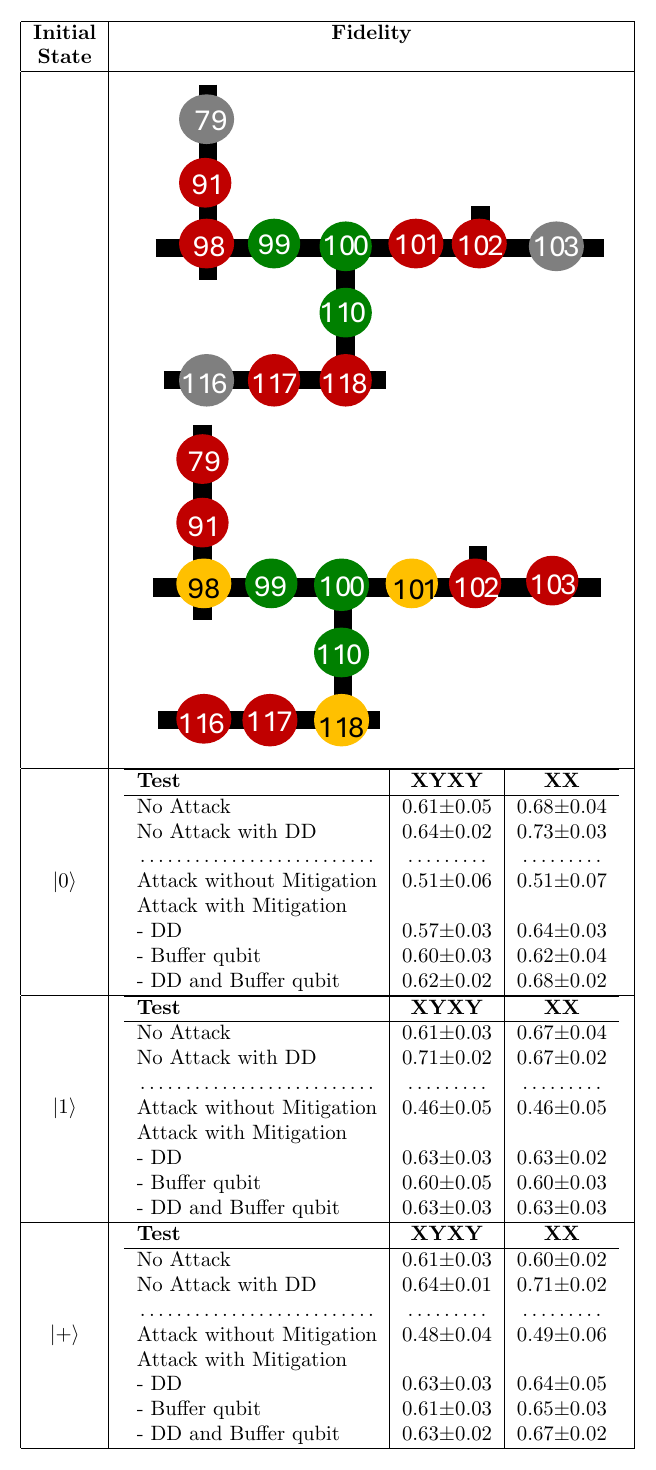}
        \caption{{\bf Average fidelity for layout (2)}.}
\label{tab:fidelity_differences_geometry_2}
\end{figure}

\begin{figure}[t!]
    \centering \includegraphics[width=0.95\columnwidth]{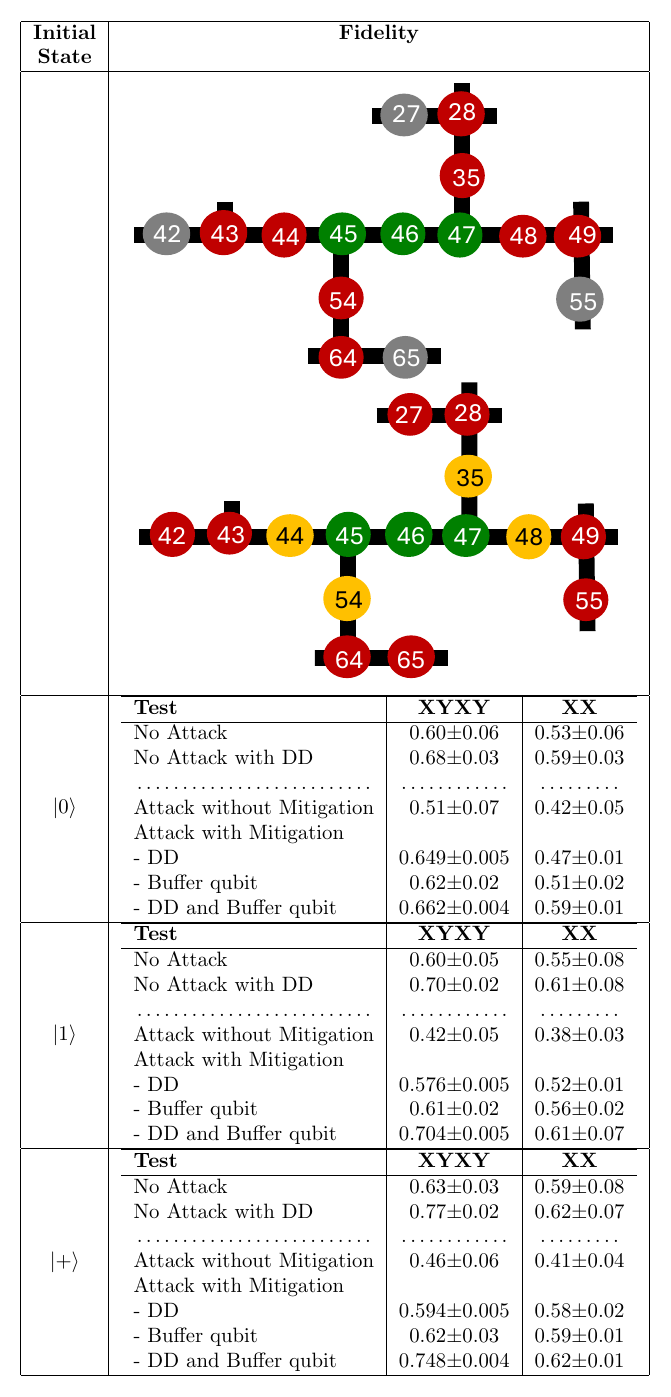}
        \caption{{\bf Average fidelity for layout (3)}.}
\label{tab:fidelity_differences_geometry_3}
\end{figure}
    
\begin{figure}[t!]
    \centering \includegraphics[width=0.95\columnwidth]{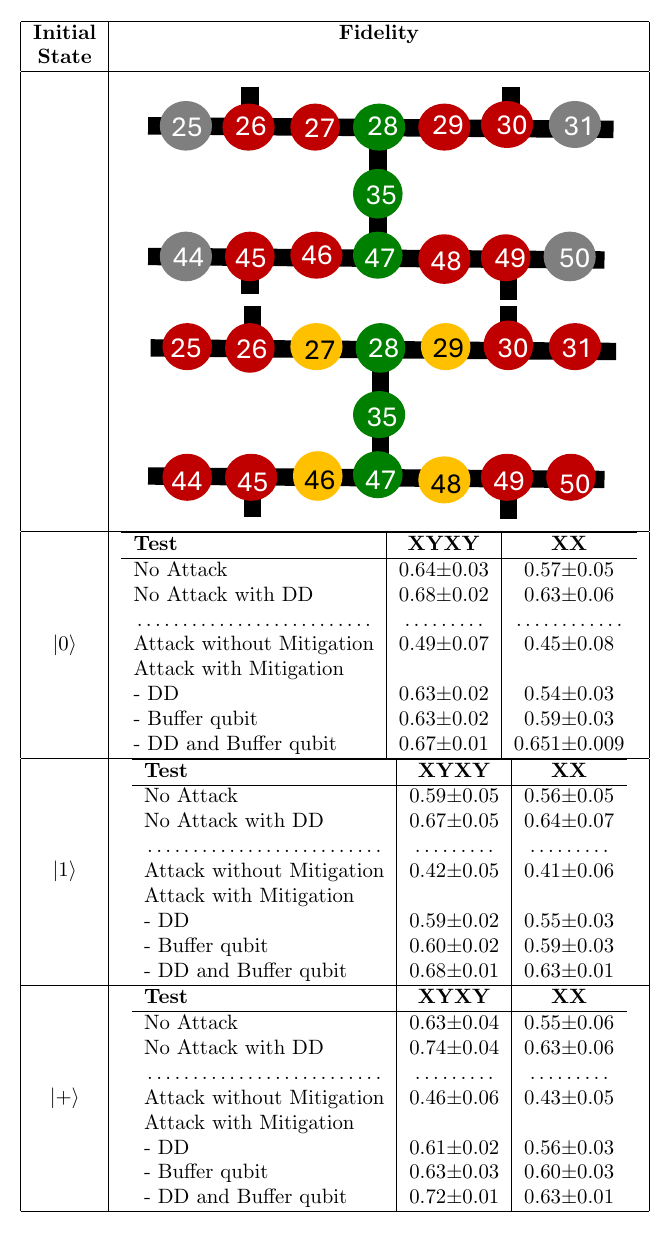}
        \caption{{\bf Average fidelity for layout (4)}.}
\label{tab:fidelity_differences_geometry_4}
\end{figure} 

\begin{figure}[t!]
    \centering \includegraphics[width=0.95\columnwidth]{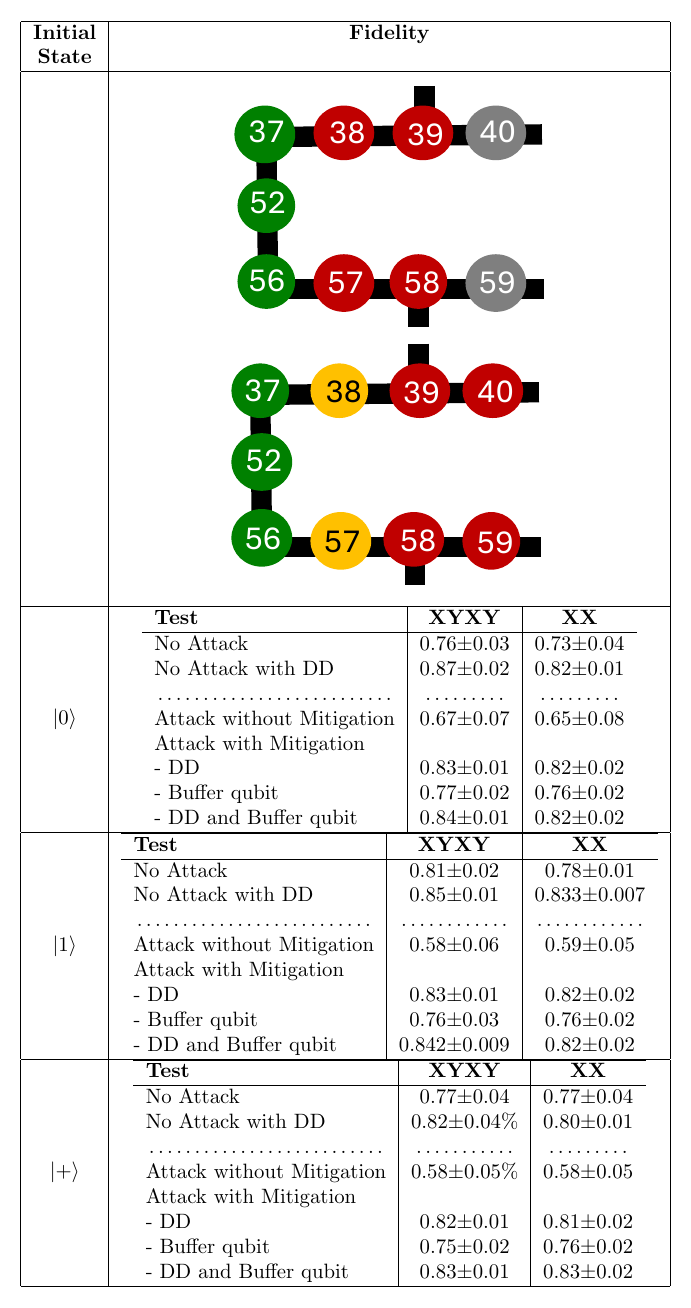}
        \caption{{\bf Average fidelity for layout (5)}.}
\label{tab:fidelity_differences_geometry_5}
\end{figure}

\clearpage
\begin{widetext}
    
\section{Quantum Hardware details}\label{app:hardware}
Qiskit quantum devices undergo daily calibration, making it essential to document all qubit descriptors to ensure result reproducibility. As shown in Table~\ref{tab:fidelity_differences_geometry_1}, the victim's circuit runs on qubits 3, 4, and 5, their performance parameters are recorded in Table~\ref{tab:qubit_performance} for 10 days of testing. Additional details for the remaining qubits can be found at~\cite{devika_git}.

\begin{table}[h]
    \centering
    
    \begin{tabular}{|c|c|}
        \hline
        \textbf{Qubit} & \textbf{Qubit performance parameters} \\
        \hline
        
        % Qubit 3 Table
        \textbf{\(Q_3\)} &
        \begin{tabular}{c|c|c|c|c|c|c|c}
            \hline
            Date & Frequency & T1 & T2 & Anharmonicity & SX Error & X Error & Readout Error \\
             & (GHz) & ($\mu$s) & ($\mu$s) & (GHz) & [\(10 ^{-3}\)] & [\(10 ^{-3}\)] & [\(10 ^{-1}\)] \\
            \hline
            12/25/2024 & 4.88 & 92.99 & 220.82 & -0.310 & 0.148 & 0.148 & 0.313 \\
            12/26/2024 & 4.87 & 261.96 & 264.66 & -0.310 & 0.208 & 0.208 & 0.308 \\
            12/28/2024 & 4.88 & 50.99 & 157.80 & -0.310 & 0.362 & 0.362 & 0.29 \\
            12/30/2024 & 4.87 & 130.64 & 133.40 & -0.310 & 0.181 & 0.181 & 0.303 \\
            01/05/2025 & 4.88 & 269.00 & 303.57 & -0.310 & 0.259 & 0.259 & 0.311 \\
            01/06/2025 & 4.88 & 273.70 & 235.40 & -0.310 & 0.199 & 0.199 & 0.353 \\
            01/07/2025 & 4.87 & 253.30 & 297.29 & -0.310 & 0.294 & 0.294 & 0.218 \\
            01/18/2025 & 4.87 & 212.87 & 220.08 & -0.310 & 0.221 & 0.221 & 0.142 \\
            01/19/2025 & 4.87 & 292.92 & 299.41 & -0.310 & 0.289 & 0.289 & 0.255 \\
            \hline
        \end{tabular} \\
        \hline
        
        % Qubit 4 Table
        \textbf{\(Q_4\)} &
        \begin{tabular}{c|c|c|c|c|c|c|c}
            \hline
            Date & Frequency & T1 & T2 & Anharmonicity & SX Error & X Error & Readout Error \\
             & (GHz) & ($\mu$s) & ($\mu$s) & (GHz) & [\(10 ^{-3}\)] & [\(10 ^{-3}\)] & [\(10 ^{-1}\)] \\
            \hline
            12/25/2024 & 4.82 & 191.99 & 96.55 & -0.310 & 0.228 & 0.228 & 0.313 \\
            12/26/2024 & 4.82 & 203.67 & 125.50 & -0.310 & 0.232 & 0.232 & 0.421 \\
            12/28/2024 & 4.88 & 276.44 & 149.43 & -0.310 & 0.275 & 0.275 & 0.04 \\
            12/30/2024 & 4.82 & 217.44 & 113.67 & -0.310 & 0.207 & 0.207 & 0.396 \\
            01/05/2025 & 4.82 & 237.58 & 137.28 & -0.310 & 0.206 & 0.206 & 0.383 \\
            01/06/2025 & 4.82 & 194.79 & 136.01 & -0.310 & 0.279 & 0.279 & 0.413 \\
            01/07/2025 & 4.82 & 176.26 & 228.67 & -0.310 & 0.219 & 0.219 & 0.184 \\
            01/18/2025 & 4.82 & 335.52 & 174.97 & -0.310 & 0.193 & 0.193 & 0.202 \\
            01/19/2025 & 4.82 & 268.35 & 312.99 & -0.310 & 0.142 & 0.142 & 0.023 \\
            \hline
        \end{tabular} \\
        \hline
        
        % Qubit 5 Table
        \textbf{\(Q_5\)} &
        \begin{tabular}{c|c|c|c|c|c|c|c}
            \hline
            Date & Frequency & T1 & T2 & Anharmonicity & SX Error & X Error & Readout Error \\
             & (GHz) & ($\mu$s) & ($\mu$s) & (GHz) & [\(10 ^{-3}\)] & [\(10 ^{-3}\)] & [\(10 ^{-1}\)] \\
            \hline
            12/25/2024 & 4.73 & 288.48 & 181.32 & -0.311 & 0.126 & 0.126 & 0.085 \\
            12/26/2024 & 4.73 & 460.19 & 228.63 & -0.311 & 0.124 & 0.124 & 0.085 \\
            12/28/2024 & 4.73 & 85.05 & 233.48 & -0.311 & 0.026 & 0.026 & 0.067 \\
            12/30/2024 & 4.73 & 233.01 & 177.18 & -0.311 & 0.248 & 0.248 & 0.086 \\
            01/05/2025 & 4.73 & 204.04 & 187.44 & -0.311 & 0.347 & 0.347 & 0.083 \\
            01/06/2025 & 4.73 & 243.73 & 180.11 & -0.311 & 0.209 & 0.209 & 0.078 \\
            01/07/2025 & 4.73 & 208.70 & 270.93 & -0.311 & 0.143 & 0.143 & 0.072 \\
            01/18/2025 & 4.73 & 84.13 & 143.00 & -0.311 & 0.028 & 0.028 & 0.075 \\
            01/19/2025 & 4.73 & 80.62 & 134.98 & -0.311 & 0.313 & 0.313 & 0.313 \\
            \hline
        \end{tabular} \\
        \hline
        
    \end{tabular}
    \caption{{\bf Qubit performance metrics over 10 days}.}
    \label{tab:qubit_performance}
    \end{table}
\end{widetext}

%\clearpage

\end{document}